\documentclass[aps,twocolumn]{revtex4-1}
\usepackage{amsmath,amssymb}
\usepackage{slashed}
\usepackage{graphicx}
\usepackage{bm}
\usepackage[T1]{fontenc}
\usepackage[utf8]{inputenc}
\usepackage{gauss} 
\usepackage[top=1.0in,bottom=1.0in,left=1.0in,right=1.0in]{geometry}
\usepackage[colorlinks,linkcolor=blue,citecolor=blue]{hyperref}
\usepackage{subfig}
\newcommand{\be}{\begin{equation}}
\newcommand{\ee}{\end{equation}}
\newcommand{\bea}{\begin{eqnarray}}
\newcommand{\eea}{\end{eqnarray}}
\newcommand{\ba}{\begin{eqnarray}}
\newcommand{\ea}{\end{eqnarray}}

\def\be{\begin{eqnarray}}
\def\ee{\end{eqnarray}}
\def\bea{\be}
\def\eea{\ee}

\def\roughly#1{\mathrel{\raise.3ex\hbox{$#1$\kern-.75em%
			\lower1ex\hbox{$\sim$}}}}

\newcommand{\dd}{\mathrm{d}}

\date{\today}
\begin{abstract}
	We analytically compute the out-of-equilibrium direct photon production rate  and electric conductivity, in a strongly coupled and  expanding Bjorken plasma,  from holography. Our results are valid at late times where the expanding plasma asymptotes Bjorken hydrodynamics. The  out-of-equilibrium rates are substantially harder and larger, early on in the Bjorken expansion phase.
\end{abstract}
\begin{document}
	\title{Out-of-equilibrium photon production and electric conductivity \\
		in a holographic Bjorken expanding plasma}
	\author{Sebastian Grieninger}
	\email{sebastian.grieninger@stonybrook.edu}
	\author{Ismail Zahed}
	\email{ismail.zahed@stonybrook.edu}
	\affiliation{Center for Nuclear Theory, Department of Physics and Astronomy,
		Stony Brook University, Stony Brook, New York 11794--3800, USA}
	\maketitle
	
	\section{Introduction}
	Thermalization in heavy-ion collisions happens ultrafast ($\sim$1-2 fm/c) producing the strongly coupled quark gluon plasma (sQGP). Once produced, the sQGP undergoes transverse and longitudinal expansion which cools it, until the eventual chemical freeze-out. The time evolution in the expansion phase is described by relativistic hydrodynamics. This evolution uses an equation of state and transport coefficients (shear and bulk viscosities) which we may be extracted from flow data. 
	
	One interesting set of observables is related to the electromagnetic emissivities (photons and dileptons),  which are emitted through the plasma and subsequent hadronic phase,  all the way to the thermal freeze-out. These direct photons and dileptons are accessible experimentally, after the
	subtraction of the emissions from the late decays in the hadronic cocktail
	(see for example~\cite{PHENIX:2014nkk,PHENIX:2012jbv,PHENIX:2008uif,ALICE:2015xmh,ALICE:2018mjj}). In contrast to the photons produced by hadronic decays in the cocktail, the direct photons -- produced in all stages -- give us valuable information about the time evolution of the produced matter in the collision, since they can escape the medium basically unaffected due to their substantially smaller  interaction. For a snapshot of the state of the theory and currently used hydrodynamic models and parameters see~\cite{Paquet:2015lta,Gale:2018ofa} (and references therein). Non-equilibrium photon emission rates and conductivities were studied in~\cite{PhysRevC.97.014901,Greif:2016jeb,Schafer:2020vvw,Yin:2013kya}.
	
	In thermal equilibrium, the  electromagnetic emission is controlled by $e^2$ at leading order in perturbation theory, and  decouples from the emitting and strongly coupled matter. Specifically,
	the photon rate is given by~\cite{Steele:1997tv,Caron-Huot:2006pee}
	\begin{equation}
	\dd\Gamma=\left.\frac{\dd^3k}{(2\pi)^3}\frac{e^2}{2|\bm{k}|}\eta^{\mu\nu}G^{<}_{\mu\nu}(k)\right|_{k^0=|\bm{k}|},
	\end{equation}
	where $k\equiv(k^0,\bm{k})$ is a null 4-vector which we put on-shell $k^0=|\bm{k}|$ and
	\begin{equation}
	G^{<}_{\mu\nu}(k)=\int\dd^4x\,e^{i (k^0t-\bm{k}\cdot\bm{x})}\langle J^\text{EM}_\mu(0)J^\text{EM}_\nu(x)\rangle\label{eq:csma}
	\end{equation}
	is the Wightman function for the electric current de-correlation. In thermal equilibrium, the Wightman function $G^{<}_{\mu\nu}$ is related 
	to the spectral density $\chi_{\mu\nu}$ using the Bose-Einstein distribution $n_b(k^0)=1/(e^{\beta k^0}-1)$
	\begin{align}
	G^{<}_{\mu\nu}(k)&=n_b(k^0)\chi_{\mu\nu}(k) \\
	&=-\frac{2}{e^{\beta k^0}-1}\text{Im}G^R_{\mu\nu}(k),
	\label{eq:wight}\end{align}
	Here $G^R_{\mu\nu}(k)$ is the retarded electric current correlator
	in Fourier space.
	
	While the emission of prompt photons can be assessed perturbatively, the photon production from a medium consisting of strongly coupled quarks and gluons in QCD is challenging~\cite{Lee:1998nz,Liu:2017fib}. The challenge becomes even greater, when the medium is out-of-equilibrium. It is therefore useful to consider this production from analogous gauge theories with gravity dual in terms of the AdS/CFT correspondence, where strong coupling calculational techniques exist. At finite temperature, supersymmetry is broken anyway, and the thermal medium densities can be used to normalize to a QCD like medium, albeit at strong coupling. Both the equilibrated electromagnetic thermal emissivities for $N=4$ super-Yang Mills (SYM) were assessed in~\cite{Caron-Huot:2006pee} and compared  to the weakly coupled emissivities from QCD, with much in sight on the role played by the strong coupling. In this spirit, we will extend the analysis to the non-equilibrium regime, where much less is known from QCD, even at weak coupling. The non-equilibrium results of our study will show that the photon equilibrium rate in~\cite{Caron-Huot:2006pee} is recovered in the long time limit, thereby providing a measure of the out-of-equilibrium effects at strong coupling. We will suggest that these effects are substantial in the photon rates in the low mass region, with possible relevance to the photon rates currently assessed at collider energies.
	
	Modeling the non-equilibrium dynamics of strongly coupled field theories from first principles is a notoriously difficult problem. In this context, holography proved to be a valuable framework to study the real-time evolution, and transport properties of certain strongly coupled field theories (see~\cite{Shuryak:2004cy,Shuryak:2008eq} for a discussion in the context of heavy-ion collisions). Within holography, the dynamics of the strongly coupled field theory is captured by general relativity in asymptotically Anti-de Sitter space. In this language, thermalization is described by the formation of a black hole~\cite{Nastase:2005rp}, whose horizon is moving away from a boundary observer~\cite{Shuryak:2005ia}. We will rely on the picture of the falling black hole, to extract the out-of-equilibrium photon production during the cooling process.
	
	Bjorken~\cite{PhysRevD.27.140} suggested a highly successful model for the central rapidity region of heavy-ion reactions,
	based on boost invariant hydrodynamics. The holographic dual gravity model which is based on the idea of the falling black hole in~~\cite{Shuryak:2005ia}, was constructed  in~\cite{Janik:2005zt,Janik:2006gp}. The main idea is to map the falling black hole onto a frame where the horizon is static and we can define (a time dependent) temperature~\cite{Janik:2006gp}. The static frame provides a well-defined framework for linear response theory. For example~\cite{Kim:2007ut,Stoffers:2011fx} used this idea to compute the diffusion of heavy quarks in these expanding backgrounds.
	
	Within holography, the equilibrium: photon and dilepton production in $\mathcal N=4$ super Yang-Mills plasma was calculated in~\cite{Caron-Huot:2006pee}. In holographic models for QCD (in the Veneziano limit) the photon production was derived in~\cite{Iatrakis:2016ugz,Iatrakis:2016yla,Iatrakis:2016qdk}. The authors in~\cite{Arefeva:2021jpa} extended the holographic discussion to anisotropic plasmas with magnetic fields. The authors of~\cite{Hassanain:2011ce} computed the plasma photoemission at strong coupling and~\cite{Mamo:2014ema} computed the gradient corrections to the photon emission rate at strong coupling. Out-of-equilibrium, the
	prompt photon and dilepton production was discussed in~\cite{Baier:2012ax,Baier:2012tc}, using the holographic model of a falling shell. However, in the context of the falling shell the background metric is not explicitly time dependent, and connects smoothly to the equilibrium case which allows the authors to rely on Fourier transformations. In this work, we will compute the out-of-equilibrium photon production in a time dependent background corresponding to a strongly coupled, Bjorken expanding plasma. Correlation functions of the Bjorken flow in the context of the holographic Schwinger-Keldysh approach were discussed in~\cite{Banerjee:2022aub}.
	
	The organization of the paper is as follows: In section~\ref{SEC_II}
	we briefly review the holographic setup for a falling black hole in bulk, dual to boost invariant Bjorken hydrodynamics on the boundary. 
	We analyze the evolution of a U(1) vector gauge field, and derive the
	on-shell boundary action from which the pertinent retarded propagator
	on the boundary can be extracted. In section~\ref{SEC_III}, we use 
	the holographic result to derive the photon emission rate in a Bjorken
	expanding and strongly coupled plasma. In section~\ref{SEC_IV} we 
	derived a closed form result for the U(1) electric conductivity in  
	out-of-equilibrium. Our conclusions are in~\ref{SEC_V}.

	\section{Holographic setup}
	\label{SEC_II}\noindent
	In the following, we study a strongly coupled SU($N_c$) $\mathcal N=4$ SYM theory at finite temperature and zero density. In order to study the photon production rate and conductivity, we couple a $U(1)$ gauge field in terms of a Maxwell term to gravity where we assume the electromagnetic coupling to be small. To be more precise, the $U(1)$ group is a subgroup of the global SU(4) $\mathcal R$-symmetry.
	
	The metric describing the asymptotic expanding fluid geometry is given by~\cite{Janik:2005zt,Janik:2006gp}.
	\begin{align}
	\!\!\dd s^2\!\!=\!\frac{R^2}{z^2}\!\!\left[-\frac{(1\!-\!v^4)^2}{1+v^4}\dd\tau^2\!\!+\!(1+v^4)(\tau^2\dd\eta^2\!\!+\dd x_\perp^2)
	\!+\!\dd z^2\!\right]\!,\label{eq:metricJP}
	\end{align}
	where $x_\perp=\{x_1,x_2\}$ are the transverse directions, $z\in\{0,1\}$ is the radial coordinate of AdS$_5$, $\tau$ is the proper time, $\eta$ the rapidity related to the longitudinal directions by $x^0=\tau\cosh\eta$ and $x_3=\tau\sinh\eta$, where $x^0$ is the time coordinate. Moreover, we set the radius of AdS and the horizon to unity. The scaling variable $v$ is given by
	\begin{equation}
	v=\frac{z}{(\tau/\tau_0)^{1/3}} \epsilon_0^{1/4}, \ \ \epsilon_0\equiv\frac14(\pi T_0)^4,
	\end{equation}
	where $\epsilon_0$ is the initial energy density and $T_0$ the initial temperature. Note that the horizon is located at $v=1$ or $z\sim \tau^{1/3}$. Following~\cite{Kim:2007ut}, we can transform the metric into a more canonical form, which resembles a static black hole. Introducing $u(z,\tau)\equiv 2v^2/(1+v^4)$ yields
	\begin{align}
	\dd s^2=&\frac{\pi^2T_0^2R^2}{u\,(\tau/\tau_0)^{2/3}}[-f(u)\dd\tau^2+\tau^2\dd \eta^2+\dd x_\perp^2]\\&
	\!\!\!+\frac{R^2}{4f(u)}\frac{\dd u^2}{u^2}+\frac{R^2}{9\,\tau^2}\dd\tau^2-\frac{R^2}{3}\frac{\tau^{-1}}{u\sqrt{f(u)}}\dd\tau\dd u,\nonumber
	\end{align}
	with $f=1-u^2$. At late times where the perfect fluid geometry is valid (i.e. $\tau\to\infty$ and $v,u=\,$const), the last two terms are suppressed, and can be ignored. Finally, rescaling the time coordinate by $t/t_0=3/2\,(\tau/\tau_0)^{2/3}$ yields
	\begin{align}
	\mathrm{d}s^2=&\frac{\pi^2 T_0^2\,R^2}{u}\!\left(\!-f(u)\frac{\tau^2_0}{t_0^2}\mathrm{d}t^2\!+\!\frac 49 t^2\frac{\tau^2_0}{t_0^2}\mathrm{d}\eta^2\!+\!\frac 32\frac{t_0}{t}\mathrm{d}x_\perp^2\!\right)\nonumber\\&+\frac{R^2}{4\,f(u)}\frac{\mathrm{d}u^2}{u^2}.\label{eq:bhstatic}
	\end{align}
	The field content of $\mathcal N=4$ SYM theory consists of $SU(N_c)$ gauge bosons, four Weyl fermions $\psi_p$, and six real scalars $\phi_{pq}$ in the adjoint representation of $SU(N_c)$. The theory has an $SU(4)$ $R$-symmetry, under which the fermions transform as $\textbf{4}$ and the scalars as $\textbf{6}$. To model electromagnetic interactions, a $U(1)$ gauge field is added to the theory, which is coupled to the conserved current of a $U(1)$ subgroup of the $R$-symmetry~\cite{Caron-Huot:2006pee}. The electromagnetic interaction is treated as being linear in the $U(1)$ gauge field for the purpose of calculating the emission rates.
	Using the background in eq.~\eqref{eq:bhstatic}, we now consider the vector perturbations $\delta a=(\delta a_t,0,\delta a_{x_\perp},\delta a_{\eta})$ of the electromagnetic $U(1)$ gauge field in radial gauge $(x_\perp=\{x_1,x_2\})$ which is captured by the bulk action
	\begin{equation}
	S_\text{matter}=-\frac{1}{4e^2}\int\dd^{5}x\sqrt{-g}\,F_{\mu\nu}F^{\mu\nu},    
	\end{equation}
	the $U(1)$ field strength tensor is  $$F_{\mu\nu}=\partial_\mu a_\nu-\partial_\nu a_\mu$$ with the $U(1)$ gauge field $a\rightarrow \delta a$. For now we will set the electromagnetic coupling $e=1$ and recover it when we compute the transport quantities. At late times, there is no dependence on the transverse directions,  and therefore we will only consider the dependence on the longitudinal direction.
	The equation of motion for the transverse fluctuations reads $a_{x_\perp}\equiv a_{x_\perp}(t,\eta,u) $\begin{widetext}
		\begin{align}
		& 4 \pi ^2 \tilde T_0^2\, u\, f(u) \left(f'(u) \,\partial_ua_{x_\perp}+f(u)\,\partial_u^2a_{x_\perp}\right)-\partial_t^2a_{x_\perp}-\frac{\partial_ta_{x_\perp}}{t}
		+\frac{9  f(u) \partial_\eta^2a_{x_\perp}}{4 t^2}=0,\label{eq:eomfull}
		\end{align}\end{widetext}
	where we defined $\tilde T_0\equiv T_0\,\tau_0/t_0$. 
	The dependence on the longitudinal direction $\eta$ is suppressed with $1/t^2$ at late times, and we can neglect it in an expansion up to order $\mathcal O(1/t)$ for large $t$. Our starting point is the geometry in eq.~\eqref{eq:metricJP}, which is only valid at late times. This justifies neglecting the $\mathcal O(1/t^2)$ contributions in the large $t$ limit we are working in. Since the last term in eq.~\eqref{eq:eomfull} is the only term containing derivatives with respect to $\eta$, we hence can drop the dependence on the longitudinal direction $a_{x_\perp}(t,\eta,u)\equiv a_{x_\perp}(t,u)$. Since the equation of motion explicitly depends on time we cannot use a simple Fourier transform, but have to perform a separation of variables by making the ansatz
	\begin{equation}
	a_{x_\perp}=c\, g_1(t)g_2(u)
	\end{equation}
	where we find that we can separate the time dependence with
	\begin{equation}
	g_1(t)=c_1 J_0\!\left(\frac{1}{2} \sqrt{\frac{3}{2}} t \sqrt{\lambda }\right)\!+\!c_2 Y_0\!\left(\frac{1}{2} \sqrt{\frac{3}{2}} t \sqrt{\lambda }\right)\!.
	\end{equation}
	Here $\lambda, c_1,$ and $c_2$ are independent of $t$ and $u$, and $J_0$ and $Y_0$ refer to Bessel functions of first and second kind, respectively. The Bessel functions are related to the Hankel functions by $J_n(t)\to \frac{1}{2} (H_n^{(1)}(z)+H_n^{(2)}(t))$ and $Y_n(t)= -\frac{1}{2} i (H_n^{(1)}(z)-H_n^{(2)}(t))$. We now set $\omega=\sqrt{3\lambda/8}$. For positive ``frequencies'' $\omega$, the solution that is ingoing at the horizon is the Hankel function of second kind $H_n^{(2)}$~\cite{Son:2002sd}. Expressing the Bessel functions in terms of the Hankel functions, and choosing the constants $c_1=1/2$ and $c_2=-i/2$, we eventually arrive at the expression
	\begin{align}
	g_1(t)&=\!\frac{1}{2}\! \left((c_1-i c_2) H_0^{(1)}\left(  \omega t\right)+(c_1+i c_2) H_0^{(2)}\left( \omega t\right)\right) \nonumber\\& =H_0^{(2)}(\omega t),
	\end{align}
	which satisfies the ingoing boundary condition at the horizon.
	We now define a Fourier-like transform using
	\begin{equation}
	a_{x_\perp}(t,u)=\int\limits_{-\infty}^\infty\frac{\dd \omega}{2\pi}\sqrt{\frac{i\,\pi\omega}{2}} H_0^{(2)}(\omega t)\psi_\omega(u)\,\tilde a_{x_\perp}(\omega),\label{eq:ft}
	\end{equation}
	where $\psi_\omega(0)=1$. This Fourier-like transformation is mathematically based on the generalized Hankel transform which we can be defined in terms of Bessel functions. Since the Hankel functions are related to the Bessel functions, we may define a Fourier-like transformation in terms of the Hankel functions of second kind. As we explain below eq.~\eqref{eq:completenessHankel} following~\cite{Kim:2007ut}, the Hankel functions do not satisfy a completeness relation, due to their singularity near zero. So restrictions on the range of validity of the transform apply. We will drop the subscript of $\psi\equiv\psi_\omega$ in the following.
	With $g_2=\psi$, we find for the spatial part
	\begin{equation}
	\frac{4u \pi^2}{\tilde T_0^{-2}}\! \left(u^2-1\right) \!\left(\left(u^2-1\right)\! 
	\psi''(u)\!+\!2 u \psi'(u)\right)+\omega ^2 \psi(u)=0\label{eq:spatial}
	\end{equation} 
	which we will solve in the following. 
	\subsection{Analytical solution}\label{secA}
	The solution to the equation of motion eq.~\eqref{eq:spatial} should behave as an ingoing wave at the horizon. Since the horizon is a regular singular point, we can expand the near-horizon solution in a power series
	\begin{equation}
	\psi(u)\sim (1-u)^\alpha(1+\ldots),
	\end{equation}
	where $\alpha=\pm\frac{i\,\omega}{4\pi\,\tilde T_0}$.
	We can recast eq.~\eqref{eq:spatial} formally as a Heun differential equation which is solved by the hypergeometric functions\begin{widetext}
		\begin{align}
		\!\!\!\! \psi_\omega(u)=& \! -i c_2\! \left(\!-\frac{1}{2}\right)^{\frac{i \omega }{2 \pi  \tilde T_0}}\! (u-1)^{\frac{i \omega }{4 \pi  \tilde T_0}} (u+1)^{\frac{\omega }{4 \pi  \tilde T_0}} u^{-\frac{\left(\frac{1}{4}+\frac{i}{4}\right) \omega }{\pi  \tilde T_0}} \, _2F_1\left(\frac{\left(\frac{1}{4}+\frac{i}{4}\right) \omega }{\pi  \tilde T_0},\frac{\left(\frac{1}{4}+\frac{i}{4}\right) \omega }{\pi  \tilde T_0}+1;\frac{i \omega }{2 \pi  \tilde T_0}+1;\frac{u-1}{2 u}\right)\nonumber\\& -i\,c_1 (u-1)^{-\frac{i \omega }{4 \pi  \tilde T_0}} (u+1)^{\frac{\omega }{4 \pi  \tilde T_0}} u^{-\frac{\left(\frac{1}{4}-\frac{i}{4}\right) \omega }{\pi  \tilde T_0}} \, _2F_1\left(\frac{\left(\frac{1}{4}-\frac{i}{4}\right) \omega }{\pi  \tilde T_0},\frac{\left(\frac{1}{4}-\frac{i}{4}\right) \omega }{\pi  \tilde T_0}+1;1-\frac{i \omega }{2 \pi  \tilde T_0};\frac{u-1}{2 u}\right).
		\end{align}\end{widetext}
	The solution satisfying the ingoing boundary condition at the horizon is the one proportional to $c_1$, which implies $c_2=0$. Demanding that $\psi_\omega(0)=1$ (since we are interested in the two point function) determines the second integration constant $c_1$, and we find\begin{widetext}
		\begin{align}
		\psi_\omega(u) =& 2^{-\frac{\left(\frac{1}{4}-\frac{i}{4}\right) \omega }{\pi  \tilde T_0}} e^{-\frac{\omega }{4 \tilde T_0}} \Gamma \left(1-\frac{\left(\frac{1}{4}+\frac{i}{4}\right) \omega }{\pi  \tilde T_0}\right) \Gamma \left(\frac{\left(\frac{1}{4}-\frac{i}{4}\right) \omega }{\pi  \tilde T_0}+1\right) (u-1)^{-\frac{i \omega }{4 \pi  \tilde T_0}} u^{-\frac{\left(\frac{1}{4}-\frac{i}{4}\right) \omega }{\pi  \tilde T_0}} (u+1)^{\frac{\omega }{4 \pi  \tilde T_0}} \,\\&\times {_2\tilde{F}_1}\left(\frac{\left(\frac{1}{4}-\frac{i}{4}\right) \omega }{\pi  \tilde T_0},\frac{\left(\frac{1}{4}-\frac{i}{4}\right) \omega }{\pi  \tilde T_0}+1;1-\frac{i \omega }{2 \pi  \tilde T_0};\frac{u-1}{2 u}\right),
		\end{align}\end{widetext}
	where $_2\tilde{F}_1(a,b;c,d)$ is the regularized hypergeometric function $_2\tilde{F}_1(a,b;c,d)/\Gamma(c)$.
	
	In our ansatz, the on-shell action gives rise to the boundary term
	\begin{align}
	& S_\text{on-shell}=\left.\frac{\pi ^4 R^5 \,\tau_0^2\, T_0^4}{4\,t_0 \,u^3}\int\dd^3x\, \dd t  A_nF^{un}\right|^{u=1}_{u=0}
	\\ &=\int\frac{\dd\omega}{2\pi}\tilde a_{x_\perp}(-\omega)\!\left[-\frac{2\pi^2}{3} \,  R \,\tilde T_0^2\, f(u) \psi_{-\omega} \psi_\omega'\!\right]^{u=1}_{u=0}\!\!\!\!\!\!\!\!\!\tilde a_{x_\perp}(\omega),\label{eq:fgreen}
	\end{align}
	where we assumed 
	\begin{equation}
	-\frac{1}{4}\int_{-\infty}^\infty\dd t\,tH_0^{(2)}(\omega t)H_0^{(2)}(-\omega't)\simeq \frac{1}{\omega}\delta(\omega-\omega'),\label{eq:completenessHankel}
	\end{equation}
	which is valid for small $\omega$. As noted above, the Hankel functions can be expressed as a combination of Bessel functions. We can establish a completeness-like relation for the Hankel transform, which is inherited from its relation to Bessel functions given by $$\int\limits_0^\infty \dd t\, t J_\nu (\omega t)J_\nu (\omega't) = \frac{1}{ \omega } \delta(\omega - \omega')\,.$$ This equation originates from the asymptotic form of a Bessel function as an exponential function over $\sqrt{t}$. However, this is not true for the Hankel function $H_{(1,2)}$ due to a singularity near zero. Despite this, using the completeness-like relation for the Hankel function is still valid for small values of $\omega$ or large times $t$, as the dominant integral contribution comes from the large time region~\cite{Kim:2007ut}. Furthermore, the perfect fluid Bjorken expanding geometry is only justified asymptotically (at late times) anyway, and the validity of our calculation is restricted to this limit. Additionally, using Hankel functions instead of Bessel functions, is necessary to match the incoming boundary condition at the black hole horizon~\cite{Son:2002sd}. To first order in $u$ the asymptotic expansion at the conformal boundary reads $$\psi\sim \psi_\textbf{(s)}+u \left(\psi_\textbf{(v)}-\frac{\psi_\textbf{(s)}\, \omega ^2 \log (u)}{4 \pi ^2 \tilde T_0^2}\right)\,.$$ To extract the expectation value, we have to subtract the divergent logarithmic contribution by adding the appropriate counterterm, but this comes at the cost of breaking conformal invariance~\cite{Horowitz:2008bn}. This means that a renormalization scale must be chosen when regulating the action. The prefactor of the logarithmic contribution enters the expectation value of the current as $$\langle J_{x_\perp}\rangle\sim\left(\psi_\textbf{(v)} -\omega ^2\psi_\textbf{(s)} /(8 \pi ^2 \tilde T_0^2)\right)\,.$$ This contact term does not affect the photon production rate, or the real part of the conductivity, since they are related to the imaginary part of the retarded Green's function. However,  it contributes to the imaginary part of the  conductivity, which is thus dependent on our choice of the renormalization scale. 
	
	From eq.~\eqref{eq:fgreen}, we can then read off the renormalized retarded Green's function $G_R(\omega)$ as\begin{widetext}
		\begin{equation}
		G^R(\omega)=\!\frac{4\pi^2}{3} R \,\tilde T_0^2\!\left[f(u) \psi_{-\omega} (u) \psi_\omega'(u) \right]_{u=0}=-\frac{R }{3}\omega \! \left(\!\omega  \left( H_{\frac{\left(\frac{1}{4}-\frac{i}{4}\right) \omega }{\pi  \tilde T_0}}\!\!+\!\psi^{(0)}\!\left(\!-\frac{\left(\frac{1}{4}+\frac{i}{4}\right) \omega }{\pi  \tilde T_0}\right)\!+\!\gamma\! +\!\log (2)\!\right)\!-\!2 \pi  \tilde T_0\!\right)\!,\label{eq:Greenbhst}
		\end{equation}
	\end{widetext}
	where $\gamma$ is the Euler number, $H(n)$
	is the $n$th harmonic number $H_n$, and $\psi^{(n)}(z)$ is the $n$th derivative of the digamma function. In general, the symmetrized Wightman function $G(\omega)$ in momentum space is related to the imaginary part of retarded Green's function via $G(\omega)=-\coth(\omega/(2\tilde T_0))$ Im$\,G_R(\omega$)~\cite{Son:2002sd}.
	On the one hand, we find for the electric conductivity at small frequencies $\omega\ll \tilde T_0$\begin{widetext}
		\begin{align}
		\sigma=\frac{1}{i\omega}G^R(\omega)
		=\frac{2\pi R}{3}\tilde T_0+\frac{1}{3}i\,R\,\log(2)\,\omega+\frac{R\,\pi\,\omega^2}{36\,\tilde T_0}+\frac{R\,\pi\,\omega^4}{4320\tilde T_0^3}+\mathcal O(\omega^5)
		\end{align}
		On the other hand, for large frequencies $\omega\gg \tilde T_0$, we find 
		\begin{align}
		\sigma=\frac{1}{i\omega}G^R(\omega)=-\frac{16\, i\, \pi ^4 R\, \tilde T_0^4}{45\,  \omega ^3}+\frac{2}{3} i R \omega \left (- \log (\pi\,  \tilde T_0)+ \log (\omega )+ \gamma -i \pi/2 -\log (2)\right)+\mathcal O\!\left(\frac{1}{\omega^{6}}\right).
		\end{align}\end{widetext}
	These results are obtained in the frame where the black hole is static and are, in general, in agreement with the conductivity of the Schwarzschild AdS$_5$ black hole in~\cite{Horowitz:2008bn} (see appendix~\ref{app1}). Note that the normalization of the metric differs compared to the Schwarzschild case.
	
	\subsection{Connecting to the boosted frame}
	So far, we worked with the frequency $\omega$ with respect to the time $t$, in the frame where the black hole is static, and given by the metric in eq.~\eqref{eq:bhstatic}. However, the time coordinate $t$ in this frame does not correspond to the proper time in the Bjorken frame since we rescaled it. Moreover, we defined our frequency with respect to this time coordinate $t$ instead of the proper time $\tau$. In order to compute the photon production rate and conductivity as seen by a physical observer, we have to convert our result to frequencies with respect to the proper time in the Bjorken frame. More specifically, this means that we need to transform  the frame where the black hole is static in eq.~\eqref{eq:bhstatic}, to the original Bjorken geometry eq.~\eqref{eq:metricJP} by considering the inverse coordinate transformation.
	We may define an inverse Fourier transform using eq.~\eqref{eq:completenessHankel},
	\begin{equation}
	G(t_1,t_2)=-\frac{1}{4}\int\limits_{-\infty}^\infty\dd \omega\,\omega\,H_0^{(2)}(\omega t_1)H_0^{(2)}(-\omega t_2)\,G(\omega).
	\end{equation}
	The same remarks regarding the range of validity discussed in section~\ref{secA},  carry to the inverse Fourier transform.
	If we introduce the relative and CM coordinates
	\begin{equation}
	s=t_1-t_2,\ {\cal T}=\frac{t_1+t_2}{2},
	\end{equation}
	we find that for $t_1,t_2\gg1$
	\begin{align}
	G&=-\frac{1}{2\pi\,\sqrt{t_1 t_2}}\int\limits_{0}^\infty \dd \omega e^{-i(t_1-t_2)\omega}G(\omega)\\&=-\frac{1}{{\cal T}}\frac{1}{\sqrt{1-\frac{s^2}{4{\cal T}^2}}}\int\frac{\dd \omega}{2\pi}e^{-i\omega s}G(\omega).\label{eq:invf}
	\end{align}
	The relative time is related to the proper time by the relation, we introduced above eq.~\eqref{eq:bhstatic} \begin{equation}
	s=3/2s_0(\tau/\tau_0)^{2/3}=\tilde \tau_0\,\tau^{2/3},\label{eq:reltime}
	\end{equation} with $\tilde \tau_0=3/2\,s_0\,\tau_0^{-2/3}$. With this in mind, we can replace the relative time $s$ in eq.~\eqref{eq:invf} by $\tau$ as defined in eq.~\eqref{eq:reltime} and find
	\begin{align}
	G&=-\frac{1}{{\cal T}}\frac{1}{\sqrt{1-\frac{s^2}{4{\cal T}^2}}}\int\frac{\dd \omega}{2\pi}e^{-i\frac{\tilde\tau_0}{\tau^{1/3}} \tau}G(\omega)\\&=-\frac{1}{{\cal T}}\frac{1}{\sqrt{1-\frac{s^2}{4{\cal T}^2}}}\int\frac{\dd q^0}{2\pi}\frac{2\tilde T_0}{3\,T(\tau)}e^{-iq^0 \tau}G\left(\frac{2\tilde T_0q^0}{3T}\right)\!,\label{eq:Greensframe}
	\end{align}
	where we introduced $$q^0=\omega\tilde\tau_0\,\tau^{-1/3},\,\,\,\,\dd\omega=\tau^{1/3}/\tilde\tau_0\,\dd q^0$$ and used $T=T_0/(\tau/\tau_0)^{1/3}$~\cite{Kim:2007ut}. Note that we introduced the proper time dependent temperature $T(\tau)$ which is related to the proper time with scaling exponent $\frac{1}{3}$~\cite{PhysRevD.27.140,Janik:2005zt}.
	
	The 4-momentum of the photon is given by 
	\begin{equation}
	q^\mu=(m_T\,\cosh(y-\eta),m_T,m_T\,\sinh(y-\eta)),
	\end{equation}
	where $y$ is the photon rapidity in the Bjorken frame with rapidity $\eta$. The number of photons per unit volume, unit rapidity and mass in the Bjorken frame, is given by
	\begin{align}
	q^0\ \frac{\dd \Gamma}{[\dd\tau\,\tau\,\dd\eta\,\dd x_\perp]\,[\dd^3q]}
	=\frac{\dd \Gamma}{\dd V_\text{Bj}\,\dd y\,\dd m_T^2/2}
	\end{align}
	It is related to the frame where the black hole horizon is fixed by 
	\begin{align}
	q^0\frac{\dd \Gamma}{[\dd\tau\,\tau\,\dd\eta\,\dd x_\perp]\,[\dd^3q]}=\omega  \frac{\dd \Gamma}{\dd V_\text{bh}\,[\dd^3k]}. \label{eq:equal}
	\end{align}
	In this subsection, we connected the frequency with respect to the time coordinate in which the black hole is static, to  the physical frequency with respect to the Bjorken frame. In particular, eq.~\eqref{eq:Greensframe} outlines how we can translate our analytical result for the Green's function eq.~\eqref{eq:Greenbhst}, to the Bjorken frame. From eq.~\eqref{eq:equal} it follows  that the number of photons per invariant spatial and phase space volume is a frame-independent quantity (since it is an experimental observable). Our result for the photon production rate, that we computed in the frame where the black hole is static (right-hand side of the equation), is thus directly related to the photon production rate in the Bjorken frame (left-hand side of the equation), which we want to compute. Combining this with our prescription to express our quantities in terms of variables in the Bjorken frame, we arrive at the main result of our paper: the out-of-equilibrium direct photon production rate to follow in eq.~\eqref{eq:photprate}.
	\section{Out-of-equilibrium direct photon production rate}
	\label{SEC_III}
\begin{figure}
		\centering
		\includegraphics[width=0.9\linewidth]{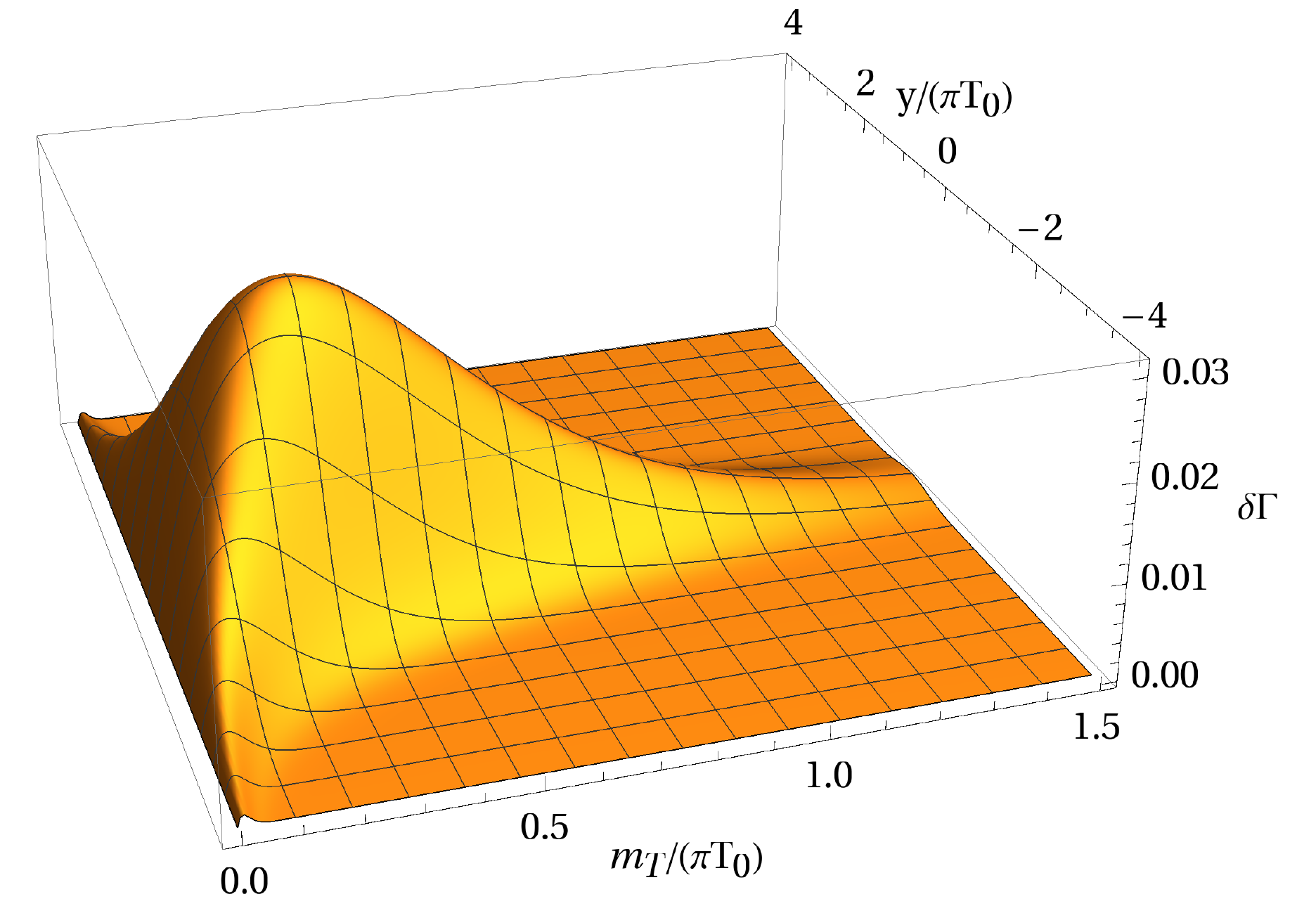}  \includegraphics[width=0.9\linewidth]{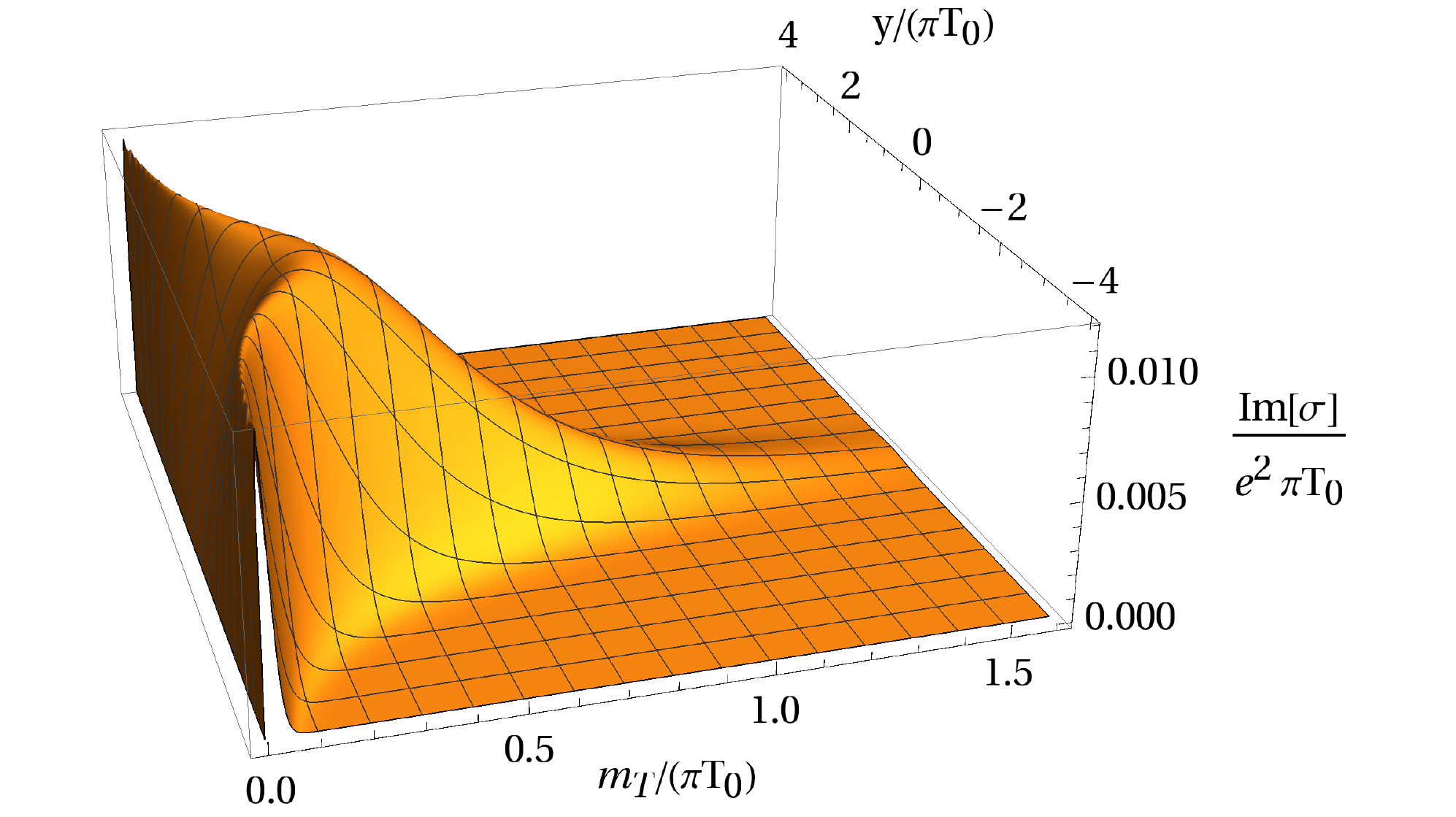}
		\caption{$\tau=5/(\pi \tilde T_0), \tilde T_0=1/\pi,\eta=0, \,\tau_0\,\pi \tilde T_0=1, R=1.$ \textit{Top:} Produced photons $\delta\Gamma\equiv\frac{m_T}{(e\pi \tilde T_0)^3}\frac{\dd\Gamma}{\dd V\dd y \dd m_T^2/2}$ as a function of the photon transverse momentum $m_T$ and rapidity $y$. For small $m_T$ and $y$ the data are proportional to the real part of the out-of-equilibrium electrical conductivity. \textit{Bottom:} The imaginary part of the out-of-equilibrium electrical conductivity. }
		\label{fig1}
	\end{figure}

\begin{figure}
		\centering
		\includegraphics[width=0.9\linewidth]{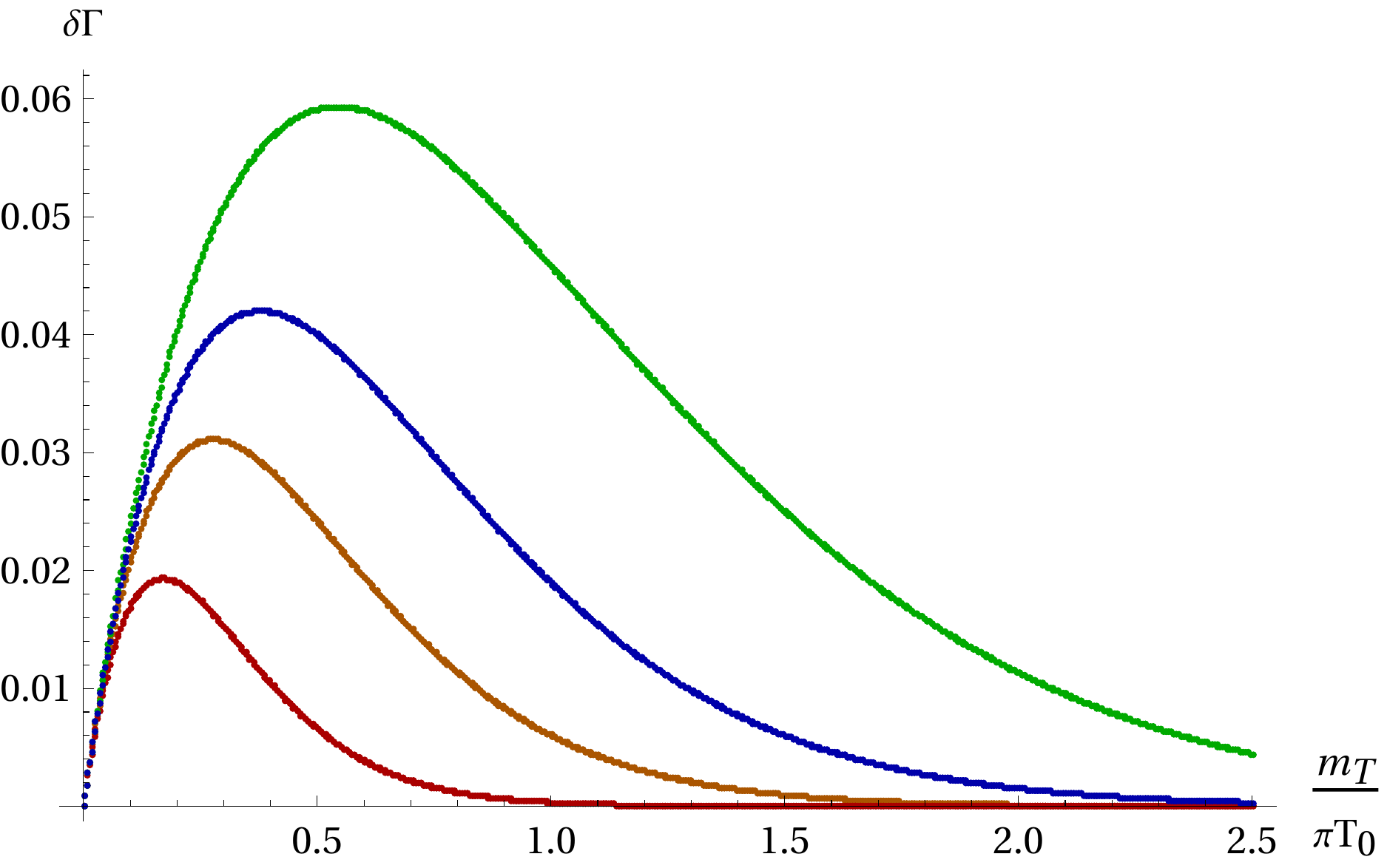}  \includegraphics[width=0.9\linewidth]{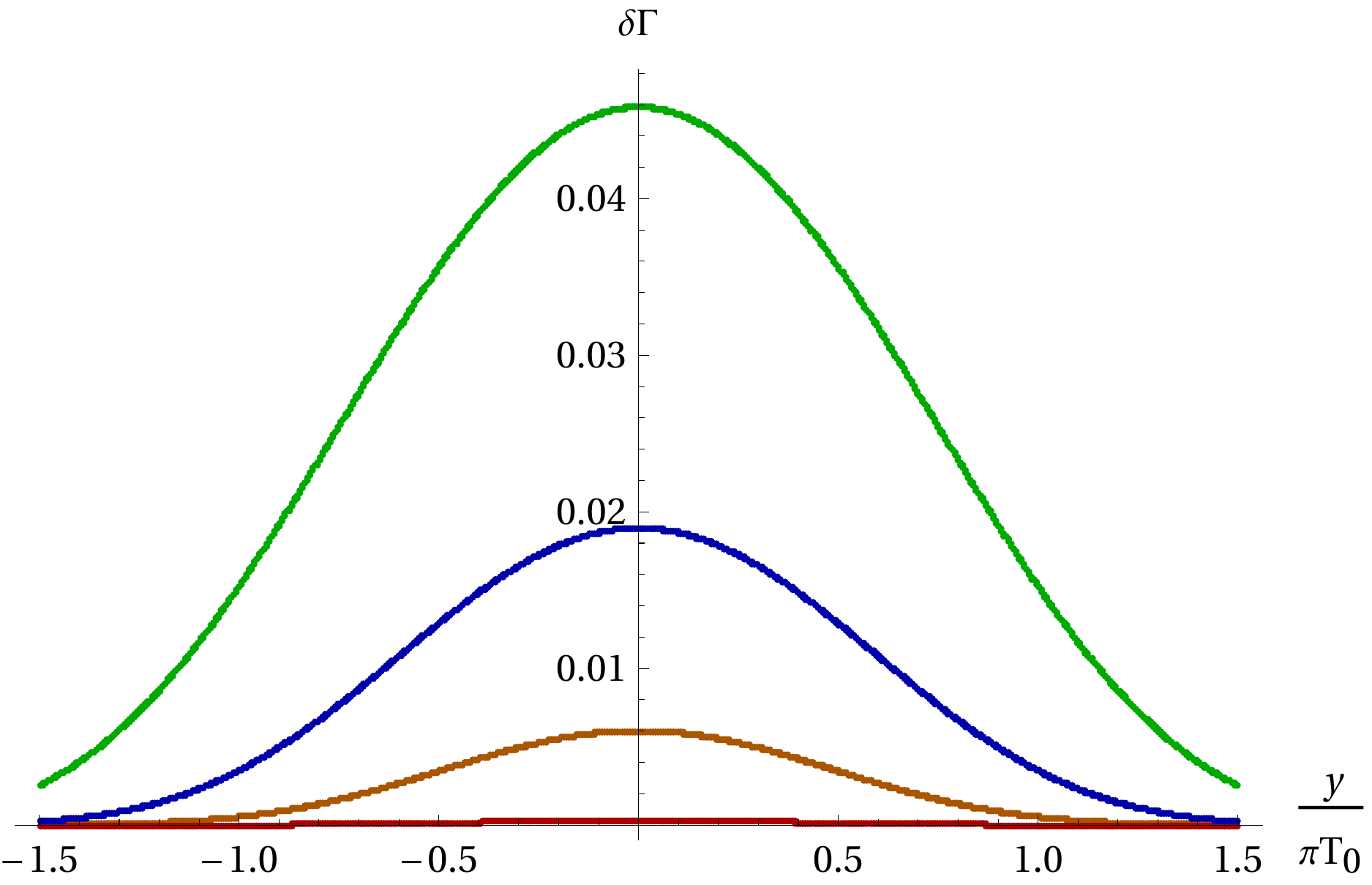}\caption{$\tau=\{2,3.25,5,10\}/(\pi \tilde T_0)$ $\text{(green, blue, brown and red curves, respectively)}, \tilde T_0=1/\pi,\eta=0, \,\tau_0 \pi \tilde T_0=1, R=1.$ \textit{Upper panel:} Produced photons $\delta\Gamma\equiv\frac{m_T}{(e\pi \tilde T_0)^3}\frac{\dd\Gamma}{\dd V\dd y \dd m_T^2/2}$ as a function of the photon momentum for different proper times indicated by different colors. We set $y=0$. \textit{Lower panel:} We set $m_T=\pi \tilde T_0$ and study the dependence on the pseudo-rapidity $y$ for different proper times.}\label{fig2}\end{figure}

	In the following, we elaborate how our solution for the retarded Green's function of the transverse gauge field fluctuations is connected to the photon production rate eq.~\eqref{eq:csma}. We can decompose the spectral function of the $\mathcal R$ current according to~\cite{Caron-Huot:2006pee}
	\begin{equation}
	G_{\mu\nu}^R(k)=P^T_{\mu\nu}(k)\Pi^T(k^0,k)+P^L_{\mu\nu}(k)\Pi^L(k^0,k),
	\end{equation}
	with the transverse and longitudinal projector $$P_{00}^T=0=P^T_{0i},\,\,\, P_{ij}^T(k)=\delta_{ij}-k_ik_j/\bm{k}^2$$  $$P_{\mu\nu}^L(k)=P_{\mu\nu}(k)-P_{\mu\nu}^T(k).$$ Taking the trace yields
	\begin{equation}
	\chi^\mu_{\ \mu}(k^0,k)=-4\text{Im}\Pi^T(k^0,k)-2\text{Im}\Pi^L(k^0,k).
	\end{equation}
	In general, the transverse and the longitudinal part contribute to the spectral function. However, we are interested in on-shell photons. For light-like momenta, the longitudinal part vanishes, and the photon-production rate is totally determined by the transverse part. Therefore,   the rate of photon production per unit rapidity and mass,
	is given by integrating over the fluid spatial evolution\begin{widetext}
		\begin{align}
		\frac{\dd \Gamma}{\dd y\,\dd m_T^2/2}&=\int\dd\tau(\tau\,\dd\eta)\,\dd x_\perp\frac{\dd \Gamma}{\dd V_\text{Bj}\dd y\,\dd m_T^2/2}\nonumber\\&=\frac{2\pi R_T^2}{{\cal T}}\int\limits_{\tau_i}^{\tau_f}\dd\tau\,\frac{1}{\sqrt{1-\frac{s^2}{4{\cal T}^2}}}\tau\int\limits_{\eta_\text{min}}^{\eta_\text{max}}\!\dd\eta\,\frac{2\,\tilde T_0 n_b(\mathfrak w(\tau,\eta))}{3\,T(\tau)}\,\text{Im}G^R_{ a_{x_\perp} a_{x_\perp}}\!\!\!\left(\mathfrak w(\tau,\eta)\right),\label{eq:photprate}\\
		\mathfrak w(\tau,\eta)&\equiv \frac{2\,\tilde T_0\,m_T}{3T}\,\cosh(y-\eta),
		\label{GTY}
		\end{align}\end{widetext}
	where $n_b$ is the Bose-Einstein distribution discussed in eq.~\eqref{eq:wight}. In figure \ref{fig1}, we display the dimensionless integrand of eq.~\eqref{eq:photprate}, multiplied by the transverse momentum $m_T$ \begin{equation}
	\delta\Gamma\equiv\frac{m_T}{(e\pi \tilde T_0)^3}\frac{\dd\Gamma}{\dd V\dd y \dd m_T^2/2}.\end{equation}
	The integrand in eq.~\eqref{eq:photprate} encodes the photon production rate per unit volume and unit rapidity, for a medium at a given proper time and rapidity $\eta$. In fig~\ref{fig1} top, we note that the photon production rate peaks in the central rapidity region, and falls off symmetrically for increasing rapidity $y$. At lower momenta $m_T$, the photon production rate stretches over a larger rapidity range, which is significantly narrow for larger momenta. The real part of the electric conductivity is related to the photon production rate displayed in the upper panel of fig~\ref{fig1}. The imaginary part of the electric conductivity  is displayed in the bottom. For very small momenta, the imaginary part of the electric conductivity vanishes in the central rapidity region. The conductivity then builds up linearly in the momentum, and peaks before falling off toward large momenta. While at very small momenta $m_T$, the  larger rapidities $y$ are the main contribution to the imaginary part of the electric conductivity, the imaginary part of the conductivity is mainly centered in the zero rapidity region after its peak. We discuss the zero rapidity region in more detail in fig~\ref{fig2} and fig~\ref{fig3}.
	If we were to integrate over the history say of a fireball, from initial to final proper time including the rapidity range, we arrive at the total number of produced photons from a medium in a given collision process.

	For $\mathfrak w\ll \tilde T_0$, we find 
	\begin{align}
	\frac{2 \tilde T_0n_b}{3T}\,\text{Im}G^R\!\!&=\frac{8 \pi  R \tilde T_0^2}{9}\label{eq:expansionrate}\\&+ \frac{ \mathfrak w \! \left(-12 T \tilde T_0^2+\mathfrak w (T^2 +2 \tilde T_0^2) \right)}{27 T^2/(\pi  R)}+\ldots.\nonumber
	\end{align}
	In order to compare with the scaling behavior of the equilibrium calculation~\cite{Caron-Huot:2006pee}, we consider the small $\mathfrak w$ expansion of the spectral density (i.e. eq.~\eqref{eq:expansionrate} without the Bose-Einstein distribution). We find that for $\mathfrak w\ll \tilde T_0$
	\begin{equation}
	\frac{2 \tilde T_0}{3T}\,\text{Im}G^R=  \frac{4 R \pi \tilde T_0^2}{9T} \mathfrak w + \frac{R \pi}{54 T} \mathfrak w^3+\mathcal O(\mathfrak w^5),
	\end{equation}
	where $\mathfrak w$ is given by eq.~\eqref{GTY}. In this late time limit, the scaling behavior reduces to the scaling behavior of the equilibrium calculation given by eq. (3.19) of~\cite{Caron-Huot:2006pee}. Our corrections are encoded in the proper time dependent temperature $T(\tau)$ and the prefactors of subleading contributions.
	
\begin{figure}
		\centering
		\includegraphics[width=0.9\linewidth]{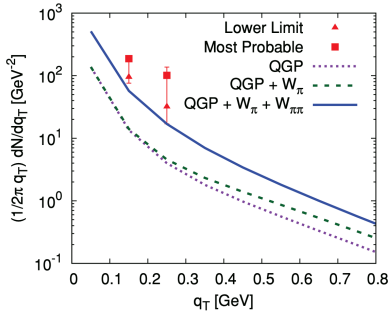}  
		\caption{Photon rates per invariant transverse momentum $m_T=q_T$ at the Super
			Proton Synchrotron (SPS)~\cite{Dusling:2009ej}. }
		\label{fig0}
	\end{figure}

	In figure~\ref{fig2}, we illustrate the dependence of the photon production on the proper time (and thus on the temperature for a given energy density). Starting from the green curve which corresponds to the smallest proper time $\tau=2$, the maximum shifts to lower momenta $m_T$ and decreases in magnitude. The red curve corresponds to $\tau=10$. As the strongly interacting medium expands and cools, the photon emissivities are reduced and shifted to lower momenta. For typically $\pi T_0\sim 1$ GeV, the shift down  is from  $\frac 34$ GeV to $\frac 14$ GeV, for a reduction in magnitude by about $\frac 12$. If we recall that for long times, our photon emissivities agree with the equilibrium rates in~\cite{Caron-Huot:2006pee} as we noted earlier, we conclude that our off-equilibrium results provide for additional enhancement of the photon emissivities at strong coupling, in relation to  weak coupling. This enhancement and down-shift of the rates in the photon intermediate and low mass region, would amount to a larger contribution stemming from a strongly coupled QGP, a welcome addition.
	Indeed the detailed analysis of the photon emissivities in~\cite{Dusling:2009ej} with their results reproduced in   Fig.~\ref{fig0}, using the weakly coupled plasma rates for the QGP, show precisely a deficit in this mass region.
	Finally, we note that in~\cite{Baier:2012ax}, the authors investigated out-of-equilibrium photon production rates in a non-expanding holographic plasma by considering a radially falling shell~\cite{Danielsson:1999zt,Danielsson:1999fa}. In contrast to our results, the absolute magnitude of the (dimensionless) photon-production rate is not monotonically falling when approaching equilibrium. However, the authors also observe that the peak is moving toward lower momenta.
	\section{Out-of-equilibrium conductivity}
	\label{SEC_IV}
	The retarded current-current correlator also contains the information about the electrical conductivity $\sigma$ of the expanding plasma, which is encoded in the zero frequency limit. More specifically, we have by~\cite{Caron-Huot:2006pee}
	\begin{equation}
	\sigma=-\left.\lim\limits_{k^0\to 0}\frac{e^2}{4i\,\tilde T_0}\frac{2}{e^{k^0/T}-1}G^R_{x_\perp x_\perp}(k)\right|_{|\bm{k}|=k^0}.
	\end{equation}
	Thus, the real part of the conductivity is given by eq.~\eqref{eq:expansionrate}. In summary, the out-of-equilibrium conductivity is given by
	\begin{align}
	\frac{\sigma}{e^2}&=\frac{ \pi  R\,T}{3}+\frac{ i R \mathfrak w  (T \log (2)+i \pi  \tilde T_0)}{6 \tilde T_0}\nonumber\\&+\frac{R\mathfrak w^2 \left(\pi  \left(T^2+2 \tilde T_0^2\right)- i T \tilde T_0 \log (64)\right)}{72  T \tilde T_0^2}.
	\end{align}

	The conductivity has dimensions of temperature.
	\begin{figure}
		\centering
		\includegraphics[width=0.9\linewidth]{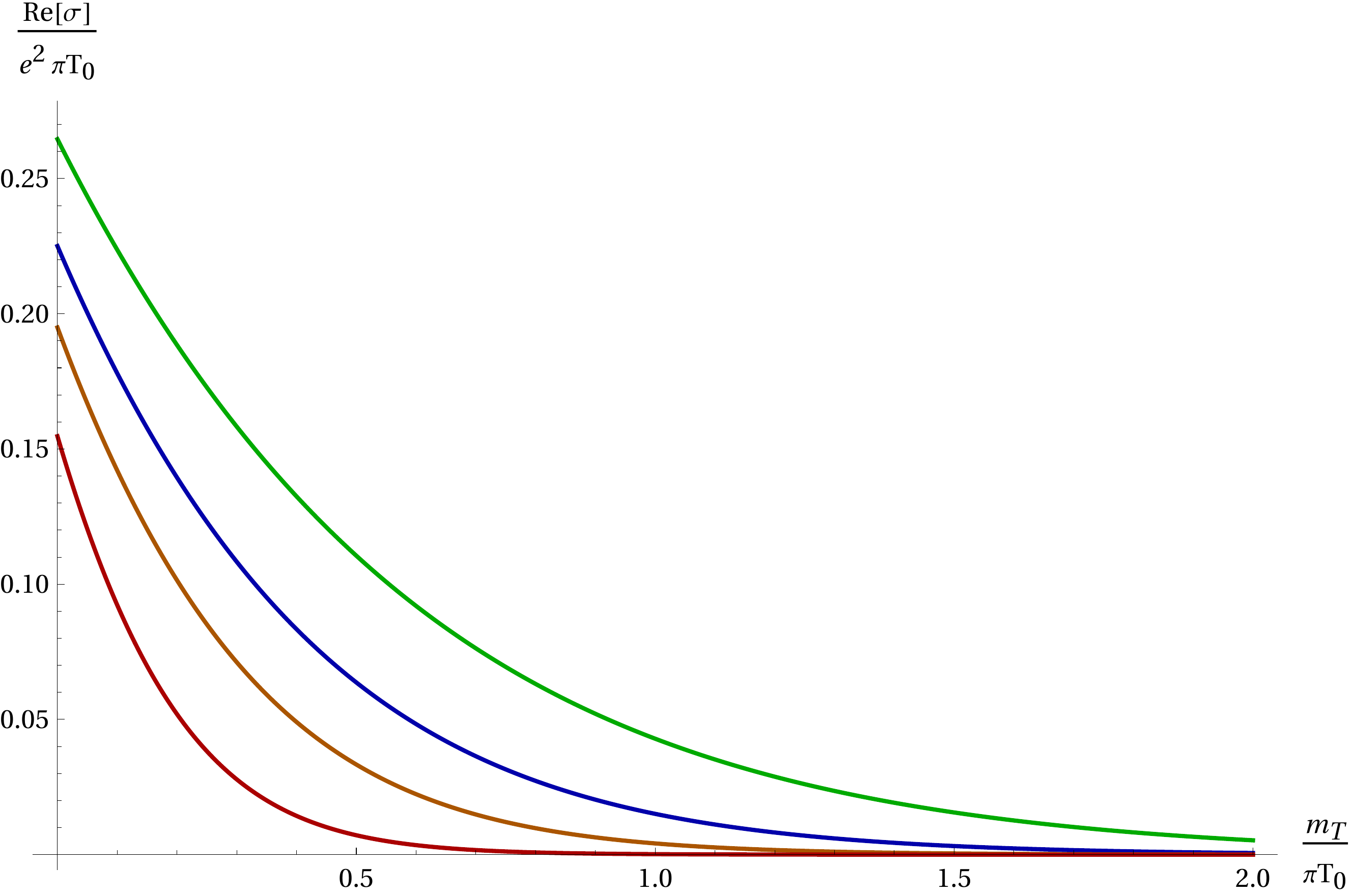}  \includegraphics[width=0.9\linewidth]{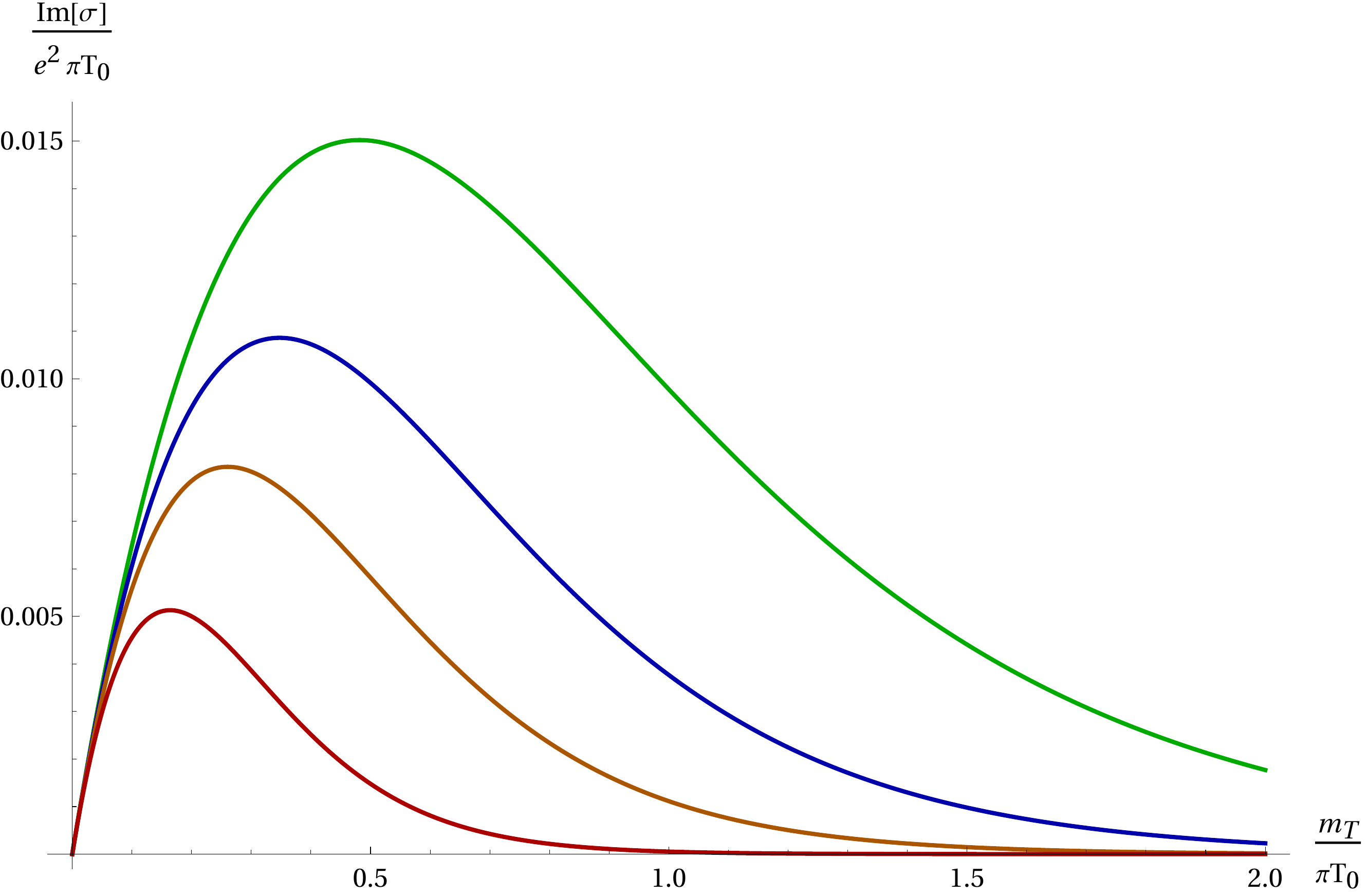}\caption{$\tau=\{2,3.25,5,10\}/(\pi \tilde T_0)$ $\text{(green, blue, brown and red curves, respectively)}, T_0=1/\pi,\eta=0, \,\tau_0\,\pi \tilde T_0=1, R=1.$ Real and imaginary part of the conductivity for different proper times and $y=0$.}\label{fig3}\end{figure}
	In figure~\ref{fig3}, we illustrate the dependence of the real and imaginary part of the conductivity on the proper time (and thus on the temperature for a given energy density). Starting from the green curve which corresponds to the smallest proper time $\tau=2$, the value of the real part at $m_T=0$ decreases for larger proper times. Furthermore, the real part drops more rapidly as a function of the photon momentum $m_T$ for increasing proper time. The maximum in the imaginary part of the conductivity,  moves toward lower frequencies and decreases in magnitude for increasing  proper time from green to red. We also note that the peak is slightly more pronounced for $\tau=10$.
	\section{Conclusions}
	\label{SEC_V}
	In this work, we derived the out-of-equilibrium direct photon production rate and electrical conductivity, for an expanding Bjorken plasma. At late times, our results agree with the literature, however, by deriving the quantities in the time dependent background our results incorporate the history of the Bjorken expansion, and are dependent on the proper time and pseudo-rapidity. Since our metric is explicitly time dependent it is not possible to rely on Fourier transforms. However, in the Bjorken limit, we were able to re-cast the metric in the form of a static black hole, and factor out the time dependence with a Fourier-like transform based on Hankel functions. This trick, which is valid for moderate frequencies, allows us to compute the out-of-equilibrium transport quantities analytically.
	
	We illustrated the dependence of the direct photon production rate on proper time, pseudo-rapidity, and photon momentum. At fixed pseudo-rapidity the peak in the production rate moves to lower momenta, for increasing proper time and is progressively suppressed. We observed a similar behavior for the imaginary part of the electrical conductivity. The real part of the conductivity for zero momenta decreases toward larger proper times as we would expect for an expanding plasma. Furthermore, it tends to zero at larger momenta.
	
	Our results provide quantitative insights into the out-of-equilibrium transport of an expanding Bjorken plasma at strong coupling. In particular, the enhancement of the photon rates in equilibrium at strong versus weak coupling noted in~\cite{Caron-Huot:2006pee}, carries to the out-of-equilibrium regime presented here. Most notably, this enhancement is
	mostly in the intermediate and low mass photon spectra. This enhancement is welcome, since current estimates using the equilibrium rates from a weakly coupled QGP plus hadrons, are still short in this mass range at the SPS energies~\cite{Dusling:2009ej}.

	It would be very interesting to extend our results to finite density along the lines of~\cite{Kalaydzhyan:2010iv,Kalaydzhyan:2011vx}, and eventually strong background magnetic fields. Another interesting direction is to study metric fluctuations in order to compute transport quantities like the shear viscosity. Moreover, it would be interesting to consider non-Abelian symmetries to compute pion yields in heavy-ion collisions with holographic techniques. Furthermore, it would be interesting to compute correction to our setup in the fluid/gravity correspondence context. In the same vein, calculation the corrections coming from the violation of the fluctuation-dissipation theorem at early times along the lines of~\cite{Caron-Huot:2011vtx,Chesler:2011ds,Chesler:2012zk} is highly interesting. We leave these tasks for future work.
	
	Finally, the out-of-equilibrium conductivity was studied in the AdS/CMT context in \cite{Bagrov:2017tqn,Bagrov:2018wzu} and it would be interesting to relate thei results to those presented here. 
	\newline\newline
	\textbf{Acknowledgements:}
	\\
	\noindent The work of S.G. and I.Z. is supported by the Office of Science, U.S. Department of Energy grant No. DE-FG88ER41450.\newline\newline

	\appendix
	\section{Conductivity in Schwarzschild AdS$_5$}\label{app1}
	The metric of the AdS$_5$ Schwarzschild black hole reads
	\begin{equation}
	\dd s^2= \frac{R^2}{u^2}\left(-f(u)\,\dd t^2+\dd\bm{x}^2+\frac{\dd u^2}{f(u)}\right).
	\end{equation}
	To compute the conductivity, we consider gauge field fluctuations about this background. The analytical solution to the gauge field equations in Fourier space at zero wave-vector and finite frequency $\omega,$ is given by~\cite{Horowitz:2008bn} and reads\begin{widetext}
		\begin{equation}
		a_x= \left(\left(\frac{1}{u}\right)^2-1\right)^{-i \omega/4} \left(\left(\frac{1}{u}\right)^2+1\right)^{-\omega /4}  {_2F_1}\left(\frac{1}{4} (-(1+i)) \omega ,1-\frac{1}{4} (1+i) \omega ;1-\frac{i \omega }{2};\frac{1}{2} \left(1-\frac{1}{u^2}\right)\right).
		\end{equation}\end{widetext}
	The renormalized retarded Green's function may be read off from
	\begin{equation}
	G^R(\omega)=-\lim\limits_{u\to 0}\frac{R\,f(u)\,a_x\,a_x'}{u},
	\end{equation}
	after subtracting the logarithmic divergence. As we noted earlier, the coefficient of the logarithm contributes  a contact term to the imaginary part of the conductivity. 
	Thus, the conductivity which is defined as $\sigma=G^R/(i\omega)$ (where we have set $T_c=\pi T$ in~\cite{Horowitz:2008bn}),  is given by\begin{widetext}
		\begin{equation}
		\sigma=-R\,\pi T+i\,\omega\,R\left[\frac12\psi\left(\frac{(1-i)\,\omega}{4\pi\,T}\right)+\frac12\psi\, \left(-\frac{(1-i)\,\omega}{4\pi\,T}\right)+\frac12\log\left(2\right)+\gamma\right],
		\end{equation}\end{widetext}
	where $\psi(u)=\Gamma'(u)/\Gamma(u)$ is the digamma function. The conductivity may be expanded in the limit of small and large frequencies compared to the temperature. The small frequency limit $\omega\ll T$ reads~\cite{Horowitz:2008bn}
	\begin{equation}
	\sigma=T\left(\pi+i\,\log\left(2\right)\frac{\omega}{2T}+\mathcal O(\omega^2)\right),
	\end{equation}
	while the large frequency limit $\omega\gg T$ is given by~\cite{Horowitz:2008bn}
	\begin{equation}
	\sigma=R\,\omega\left(\frac\pi2+i(\log\frac{\omega}{2\pi\,T}+\gamma)+\mathcal O(\omega^{-4})\right).
	\end{equation}
	\bibliography{references}
\end{document}